\documentclass{nature}
\usepackage{amssymb,amsmath}
\usepackage{graphicx}
\usepackage{hyperref}
\usepackage{array}
\usepackage{url}

\title{Unexpected features of branched flow through high-mobility two-dimensional electron gases} 

\author{M.P. Jura$^{1*}$, M.A. Topinka$^{2,3*}$, L. Urban$^4$, A. Yazdani$^4$, H. Shtrikman$^5$, L.N. Pfeiffer$^6$, K.W. West$^6$, \& D. Goldhaber-Gordon$^{2\dag}$}

\begin{document}

\maketitle

\begin{affiliations}
 \item Department of Applied Physics, Stanford University, Stanford, California 94305, USA
 \item Department of Physics, Stanford University, Stanford, California 94305, USA
 \item Department of Material Science \& Engineering, Stanford University, Stanford, California 94305, USA
 \item Department of Physics, Princeton University, Princeton, New Jersey 08544, USA
 \item Department of Condensed Matter Physics, Weizmann Institute of Science, Rehovot 76100, Israel
 \item Bell Labs, Alcatel-Lucent, Murray Hill, New Jersey 07974, USA

* These authors contributed equally to this work

\dag e-mail: goldhaber-gordon@stanford.edu
\end{affiliations}

\noindent
Published as:
\newline\noindent
Jura, M.~P., Topinka, M.~A., Urban, L., Yazdani, A., Shtrikman, H., Pfeiffer, L.~N., West, K.~W. \& Goldhaber-Gordon, D.  Unexpected features of branched flow through high-mobility two-dimensional electron gases. \textit{Nature Physics} \textbf{3}, 841-845 (2007).
\newline\noindent 
\url{http://www.nature.com/nphys/journal/v3/n12/abs/nphys756.html}

\newpage
\begin{abstract}
GaAs-based two-dimensional electron gases (2DEGs) show a wealth of remarkable electronic states\cite{FQHE,StripeState,ZRS}, and serve as the basis for fast transistors, research on electrons in nanostructures\cite{Sohn-Review,vanderWiel-Review}, and prototypes of quantum-computing schemes\cite{Petta-Science}. All these uses depend on the extremely low levels of disorder in GaAs 2DEGs, with low-temperature mean free paths ranging from microns to hundreds of microns\cite{100micron}.  Here we study how disorder affects the spatial structure of electron transport by imaging electron flow in three different GaAs/AlGaAs 2DEGs, whose mobilities range over an order of magnitude.  As expected, electrons flow along narrow branches that we find remain straight over a distance roughly proportional to the mean free path.  We also observe two unanticipated phenomena in high-mobility samples.  In our highest-mobility sample we observe an almost complete absence of sharp impurity or defect scattering, indicated by the complete suppression of quantum coherent interference fringes.  Also, branched flow through the chaotic potential of a high-mobility sample remains stable to significant changes to the initial conditions of injected electrons.
\end{abstract}

Scanning gate microscopy (SGM) images of electron flow in two-dimensional electron gases (2DEGs)\cite{Eriksson-SGM,Topinka-Science,Topinka-Nature,LeRoy-LocalDensity,Topinka-PhysicaE,LeRoy-prism,Topinka-PhysicsToday,LeRoy-Interferometer,Crook-Collimation,Crook-Cyclotron, Ensslin-QD} provide direct spatial information not available in conventional electrical transport measurements.  Our SGM studies show how varying disorder affects electron flow, and enable us to infer information about the disorder potential in our different samples.  Achieving a detailed picture of the disorder potential\cite{Yacoby-SET,Steele-PRL} may help to understand why exotic electron organization emerges in some 2DEGs and not others, and to aim for ever weaker disorder or even tailored disorder\cite{Pepper-SpinLattice}.  By analyzing the differences between images of flow in our samples, we find that the higher-mobility samples are increasingly dominated by small-angle scattering instead of hard-scattering\cite{Coleridge}.  Finally, we investigate an unusual property of electron flow through such a small-angle scattering disorder potential: though the disorder potential is classically chaotic, branches of flow are stable to significant changes in initial conditions.

On each of three 2DEG samples defined in GaAs/AlGaAs heterostructures (Table~\ref{proptable}), we use a home-built scanning gate microscope to image the flow of electrons emanating from a split-gate quantum point contact (QPC)\cite{vanWees,Wharam} at $4.2 \ \mathrm{K}$, as schematically shown in Fig. 1d.  Using a recently established technique\cite{Topinka-Science,Topinka-Nature,LeRoy-LocalDensity,Topinka-PhysicaE,LeRoy-prism,Topinka-PhysicsToday,LeRoy-Interferometer}, we measure the conductance across the QPC while scanning a sharp conducting tip ${\sim}20 \ \mathrm{nm}$ above the surface of the sample.  We negatively bias the tip to create a depletion region in the 2DEG below.  When the tip is above a region of high electron flow from the QPC, it backscatters electrons through the QPC, reducing the measured conductance.  By scanning the tip and recording the drop in conductance $\Delta G$ for each tip location, we thus image electron flow.  Images of flow gathered in this way have been found to accurately reproduce the underlying, unperturbed flow patterns\cite{Topinka-Nature,Shaw-Thesis}.  Unless otherwise stated, the conductance of the QPC is set to the first plateau, $2e^2/h$.  Samples with mobilities as high as samples B and C or as low as sample A have not been previously imaged using this technique.

Even in relatively clean 2DEGs, in which the mean free path is much greater than the Fermi wavelength, electrons do not propagate smoothly.  Instead they often flow along caustics -- narrow branches of concentrated current density -- which form owing to the cumulative effect of passing over many small-angle scattering sites that focus electrons into these branches\cite{Topinka-Nature,Shaw-Thesis,Heller-Branching}.  The disorder potential is composed of a low, bumpy background of small-angle scattering sites from ionization of some of the Si donors present above any conventional GaAs-based 2DEG, as well as sharp, tall spikes, causing hard-scattering, from unintentional impurity atoms embedded in the 2DEG, more prevalent in lower-mobility 2DEGs\cite{Coleridge}.  The length scale of potential fluctuations in the 2DEG produced by ionized donors in the donor layer is set by the separation between the 2DEG and the donor layer, a distance which is significantly shorter than the mean free path.

Figure 1 shows the striking differences in flow patterns for electrons traveling in 2DEGs with a wide range of mobility and mean free path.  The flow varies from twisted and diffusive in the lowest-mobility sample (sample A) to straight, smooth branches of flow in the highest-mobility sample (sample C).  Here we measure that the characteristic length $l_B$ over which branches remain straight increases in samples with longer mean free path $l$, as might be expected.  As a simple way to estimate $l_B$, we average the distances along branches between all observable pairs of points where one branch splits into two, in five images each for samples B and C (see Supplementary Information, Section III); we identify $49$ and $29$ branch points in all images for samples B and C, respectively.  We find in sample B with $l = 13 \ \mathrm{\mu m}$ that $l_B = 480 \ \mathrm{nm}$, and in sample C with $l = 28 \ \mathrm{\mu m}$ that $l_B = 740 \ \mathrm{nm}$.

We now use SGM to estimate the density of hard-scatterers in our highest-mobility samples, something which is difficult to do from basic transport measurements.  In Fig. 2, we discover an almost complete absence of hard-scattering from impurities and defects in the highest-mobility material by comparing interference fringes which decorate images of flow in the three samples.  All previous sufficiently high-resolution images of electron flow through QPCs using this backscattering imaging technique have shown interference fringes\cite{Topinka-Science,Topinka-Nature,LeRoy-LocalDensity,Topinka-PhysicaE,LeRoy-prism,Topinka-PhysicsToday,LeRoy-Interferometer}.   However, as shown in Fig. 2, in our highest-mobility sample, sample C, the interference pattern is completely suppressed.  Sample A, with the lowest mobility, shows strong interference fringes throughout the flow pattern, while sample B, with intermediate mobility, shows interference fringes only in a few locations.

As schematically shown in insets to Fig. 2, and as detailed in previous work\cite{Topinka-Nature,LeRoy-Interferometer,Shaw-Thesis,Heller-ThermalWP}, our imaging technique yields fringes due to the interference of two sets of paths- path 1 (blue), the roundtrip path electrons take from the QPC to the tip and back, and paths 2 (red), the sum of all roundtrip paths electrons take from the QPC to all fixed scatterers in the sample and back.  When we move the tip away from the QPC by half the Fermi wavelength,  $\lambda_F/2$, the phases of paths 2 do not change, but the phase accumulated along path 1 increases by $2\pi$; thus, we expect a fringe spacing of $\lambda_F/2$.  As has been shown previously\cite{Topinka-Nature}, the distance from the QPC over which these fringes persist is not cut off by thermal broadening (see Supplementary Information, Section VI) and can be expected to be as long as the quantum coherence length, which is $9 \ \mathrm{\mu m}$  for sample B and $6 \ \mathrm{\mu m}$ for sample C at $4.2 \ \mathrm{K}$\cite{Giuliani}.

In high-mobility sample C (Fig. 2c), we observe no interference fringes because the disorder potential contains very few sites that hard backscatter electrons back through the QPC: paths 2 apparently have much smaller amplitude than path 1.  In sample B, interference fringes are only visible in a few rare locations, one of which is shown in Fig. 2b, signifying the presence of a hard-scatterer near the center of interference fringes.  The rare appearance of fringes is consistent with this high-mobility sample containing a sparse sprinkling of impurities close to the 2DEG.  Judging from the density of observed scattering centers in the flow through five devices in sample B (see Supplementary Information, Section IV), we roughly calculate the density of hard-scatterers is $\sim 3/\mathrm{\mu m}^2$, which is close to estimates in similar materials\cite{Coleridge}.  In sample C, the lack of any observable fringes indicates that the density of hard-scatterers is less than $\sim 0.5/\mathrm{\mu m}^2$.  Sample A contains too many scattering sites to establish a reliable estimate.

Figure 3 shows an interesting consequence of small-angle scattering in a high-mobility sample (sample B): the branched flow pattern is robust to changes in initial conditions.  Previous experiments have inferred information about electron trajectories in quantum chaotic systems from transport measurements\cite{Marcus-QC}, but here we use SGM to directly image these trajectories and their sensitivity to initial conditions.  Building on previous experiments\cite{Topinka-PhysicaE,LeRoy-prism}, we change the initial position and angle at which electrons are launched into the 2DEG by applying unequal potentials to the two QPC gates.  By putting as much as a $500 \ \mathrm{mV}$ offset between the two gate voltages, simulations indicate (see Methods section) that we can shift the position of injected electrons by $\pm 30 \ \mathrm{nm}$, and the initial injection angle by $\pm 5^{\circ}$, away from the more negatively biased gate.  A $60 \ \mathrm{nm}$ shift is significant in this system because it is similar to both the $50 \ \mathrm{nm}$ correlation length of our simulated disorder potential and the ${\sim}65 \ \mathrm{nm}$ width we calculate for the QPC channel.  

Electron trajectories in a low, bumpy disorder potential are classically chaotic: very large changes in the flow pattern might be expected from this significant change in initial conditions.  Contrary to these expectations, the experimental flow pattern appears to be remarkably immune to changes in electron injection position and angle.  As we shift the QPC potential, electrons flow along the same set of branches, with the only difference being changes in the relative strength of flow along each branch.  Figure 3a-c shows experimental images for which the gates are unevenly charged and electrons from the first conductance plateau are pushed to the left, not pushed and pushed to the right, respectively.  Comparing the images in Fig. 3a-c, it is remarkable that the central branches $4 \ \mathrm{\mu m}$ away from the center of the QPC (visible in Fig. 3a-d, denoted with brown arrows) deviate by less than $10 \ \mathrm{nm}$, despite our estimate that the center of the QPC moves $60 \ \mathrm{nm}$.  Figure 3d shows electron flow when the QPC is symmetrically opened to the second conductance plateau, $4e^2/h$, so that a wider spread of electrons emanates (through both the first and second conductance modes).  The positions of the left branches (visible in Fig. 3a,d, denoted with a blue arrow) and right branches (visible in Fig. 3c,d, denoted with a green arrow) are extremely stable as well.

To understand these data, in Fig. 4 we numerically simulate flow in a low, bumpy disorder potential chosen to match the physical characteristics (Fermi energy $7.5 \ \mathrm{meV}$, mean free path $13 \ \mathrm{\mu m}$, and donor setback $25 \ \mathrm{nm}$) of sample B  (see the Methods section).  In Fig. 4a,b, we show the chaotic nature of the potential by displaying the extreme sensitivity of individual classical trajectories (shown in red) to small changes in initial conditions.  We calculate that the Lyapunov length, the average length scale over which two trajectories initially separated by a small displacement exponentially diverge, is $300 \ \mathrm{nm}$ for this potential (see Supplementary Information, Section VIII).  Based on these results, we might expect branch shapes and positions on the scale of these images to be highly sensitive to initial conditions.  Contrary to this expectation, the overall simulated branched flow pattern (shown in blue) in Fig. 4a and Fig. 4b, composed of an ensemble of many classical trajectories with an angular spread similar to that expected from a QPC, demonstrates remarkable immunity to changes in initial conditions.  This stability is in accord with previous theoretical work on stability in chaotic systems \cite{Berry-Stability,Tomsovic-Stability,Heller-Stability} which found that while individual trajectories are unstable the overall flow pattern, or manifold, can be much more stable to variations in initial conditions.

However, whereas classical mechanics has previously been found adequate to explain branch formation\cite{Topinka-Nature} and here reproduces stability of branches to small changes in initial position, it cannot reproduce the full experimentally observed stability of flow patterns.  The simulated classical flow pattern in Fig. 4c, which is shifted by $60 \ \mathrm{nm}$, is very different from that in Fig. 4a, demonstrating that the branches in classical flow are not stable to a full $60 \ \mathrm{nm}$ shift in initial position of a point source (see Supplementary Information, Section II for more information).  We find that a quantum treatment is necessary to reproduce the observed experimental robustness, as in Fig. 4d,e.  The positions of most of the branches in the simulated quantum flow pattern deviate by less than $10 \ \mathrm{nm}$ even several microns away from the QPC, even though there is less than $5 \%$ overlap between the injected electron $|\psi|^2$ in the two simulations (see inset Fig. 4e).  We posit that not only diffraction out of the QPC but also quantum scattering off the disorder potential steadily increases the spatial overlap between the two differently injected beams as they move away from the QPC, providing an additional source of stability in the quantum simulations.

In conclusion, the technique we describe here, which enables our probe to act as a detector for individual hard-scattering sites, could prove useful as a measurement of the impurity concentration in the highest-mobility 2DEGs, and could provide important insight into the limiting factor of the ultimate mobility in GaAs-based 2DEGs.  Also, the branch stability we observe could be important in applications which require guiding electrons through 2DEGs and related low-dimensional semiconductor structures.

\begin{methods}
To determine the potential produced at the 2DEG by gates on the surface of the sample, we use SETE, a three-dimensional Poisson-Schr\"{o}dinger solver written by M. Stopa (Director of National Nanotechnology Infrastructure Network at Harvard University).  SETE first determines the profile of the conduction band edge along the growth direction of the heterostructure by solving the Schr\"{o}dinger equation in one dimension and then calculates the potential and electron density at the 2DEG self-consistently from the Poisson equation\cite{Stopa-SETE}.  We then quantum-mechanically time-evolve the first mode of conductance through the QPC potential computed by SETE.  Based on the resulting quantum-mechanical wave function we estimate how much we shift the initial position and angle of injected electrons by unevenly charging the QPC gates.  In addition, we determine an appropriate spread of angles to use with our classical trajectories: the full-width at half-maximum of electron trajectories emanates at $\pm 10 ^{\circ}$.  

To simulate electron flow in sample B, we create the disorder potential from a simple physical model based on two experimentally known parameters: the distance between the 2DEG and donor layer, and the mean free path (see Supplementary Information, Section VII, for more information).  We consider only donors in the donor layer, and assume each ionized donor is metallically screened by the 2DEG, creating a density perturbation corresponding to a potential dip with a full-width at half-maximum of $40 \ \mathrm{nm}$.  We convolve a random donor distribution with this potential.  This results in a disorder potential with correlation length $50 \ \mathrm{nm}$.  We then set the amplitude of the potential energy fluctuations so that the quantum-mechanically simulated mean free path in this disorder potential matches the measured value in our real sample.  We use this disorder potential in our classical simulations, and in the quantum-mechanical simulations we add it to the gate potential determined by SETE.  In the quantum-mechanical simulations, we then quantum-mechanically time-evolve a wavepacket with energy spread equal to $3.5 \ kT$ to account for the effects of thermal averaging\cite{Heller-ThermalWP}.
\end{methods}

\begin{addendum}
\item[Acknowledgements]We thank A. Sciambi for help characterizing the 2DEG samples and determining conditions for stable gating.  We are grateful to E.~J. Heller, G.~A. Fiete and Y. Imry for discussions about theory and suggestions about many aspects of the work.  We thank M. Stopa for advice on using SETE as well as NNIN/C for making SETE available.  This work was supported by the Stanford-IBM Center for Probing the Nanoscale, an NSF NSEC, grant PHY-0425897.  Work was performed in part at the Stanford Nanofabrication Facility of NNIN supported by NSF, grant ECS-9731293.  M.P.J. acknowledges support from a Stanford Graduate Fellowship during the beginning of this work and a NDSEG Fellowship.  M.A.T. recognizes support from an Urbanke Fellowship.  D.G.-G. recognizes support from a Packard Fellowship.
\item[Correspondence]Correspondence and requests for materials should be addressed to D.G.-G.
\item[Supplementary Information]Supplementary information can be found online at 
\newline\noindent
\url{http://www.nature.com/nphys/journal/v3/n12/abs/nphys756.html}
\item[Competing financial interests] The authors declare that they have no competing financial interests.
\end{addendum}

\subsection{References}

\newcommand{\noopsort}[1]{} \newcommand{\printfirst}[2]{#1}
  \newcommand{\singleletter}[1]{#1} \newcommand{\switchargs}[2]{#2#1}

\newpage

\begin{table}
\begin{tabular}{|c|c|c|c|}
\hline Sample & A & B & C\\
\hline density, $n \ (10^{11} \, \mathrm{cm}^{-2})$ & $4.3$ & $2.1$ & $1.5$\\
\hline mobility, $\mu \ (10^6 \, \mathrm{cm}^2/\mathrm{V\,s})$ & $0.14$ & $1.7$ & $4.4$\\
\hline mean free path, $l \ (\mathrm{\mu m})$ & $1.5$ & $13$ & $28$\\
\hline distance from surface to 2DEG $(\mathrm{nm})$ & $57$ & $68$ & $100$\\
\hline distance from donors to 2DEG $(\mathrm{nm})$ & $22$ & $25$ & $68$\\
\hline
\end{tabular}
\caption[Properties of three 2DEG samples]{\label{proptable}\footnotesize{Properties of 2DEG samples, measured in a Hall bar configuration.  Sample A was grown by the commercial grower IQE.  Sample B was grown by H. Shtrikman.  Sample C was grown by L. N. Pfeiffer and K. W. West.  Each of the three heterostructures hosts a 2DEG at a single GaAs/AlGaAs heterojunction, not in a square quantum well.}}
\end{table}

\newpage

\begin{figure}
\begin{center}
\includegraphics[width=6.5in]{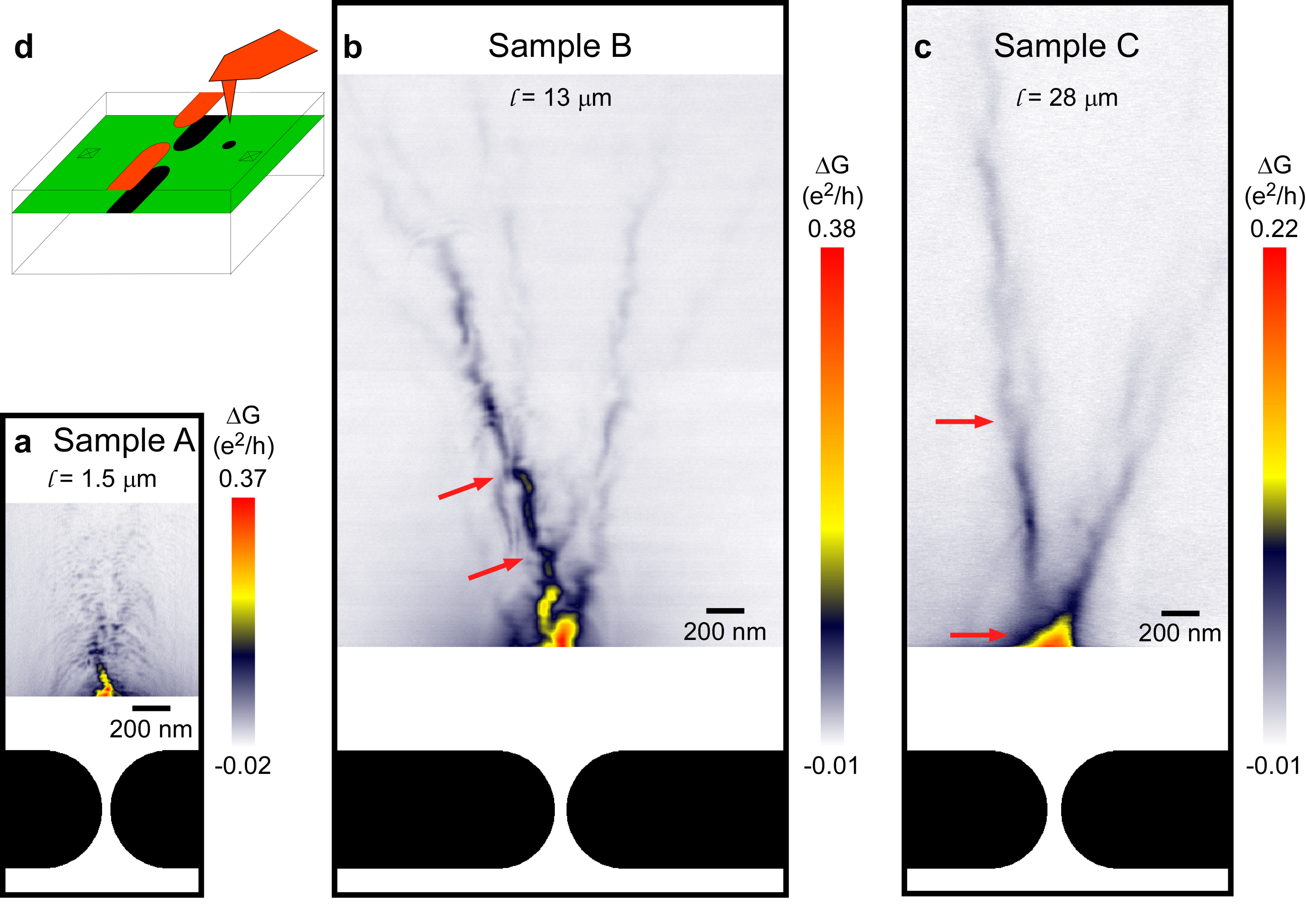}
\end{center}
\caption[Branches in three samples]{\label{branchfig}\footnotesize{\textbf{Electron flow in samples with different mobilities.}  \textbf{a}-\textbf{c}, Electron flow originates from the QPC formed between the two depletion regions, schematically indicated in black below each image.  The length over which branches remain straight depends on the sample's mean free path.  \textbf{a}, Sample A, with mean free path $1.5 \ \mathrm{\mu m}$, has branches that quickly change direction or are hard to assign a direction.  \textbf{b}, Sample B, with mean free path $13 \ \mathrm{\mu m}$, has branches that remain straight over moderate length scales.  The average distance along a branch of flow between observable branch points is $480 \ \mathrm{nm}$.  \textbf{c}, Sample C, with mean free path $28 \ \mathrm{\mu m}$, has branches that remain straight over the longest length scales.  The average distance along a branch of flow between branch points is $740 \ \mathrm{nm}$.  \textbf{b},\textbf{c},  Red arrows denote two neighboring, observable branch points where one branch splits into two.   \textbf{d}, Schematic showing metallic gates and tip (orange) creating depletion regions (black) in the 2DEG (green) buried below the surface of the sample.}}
\end{figure}

\newpage

\begin{figure}
\begin{center}
\includegraphics[width=6.5in]{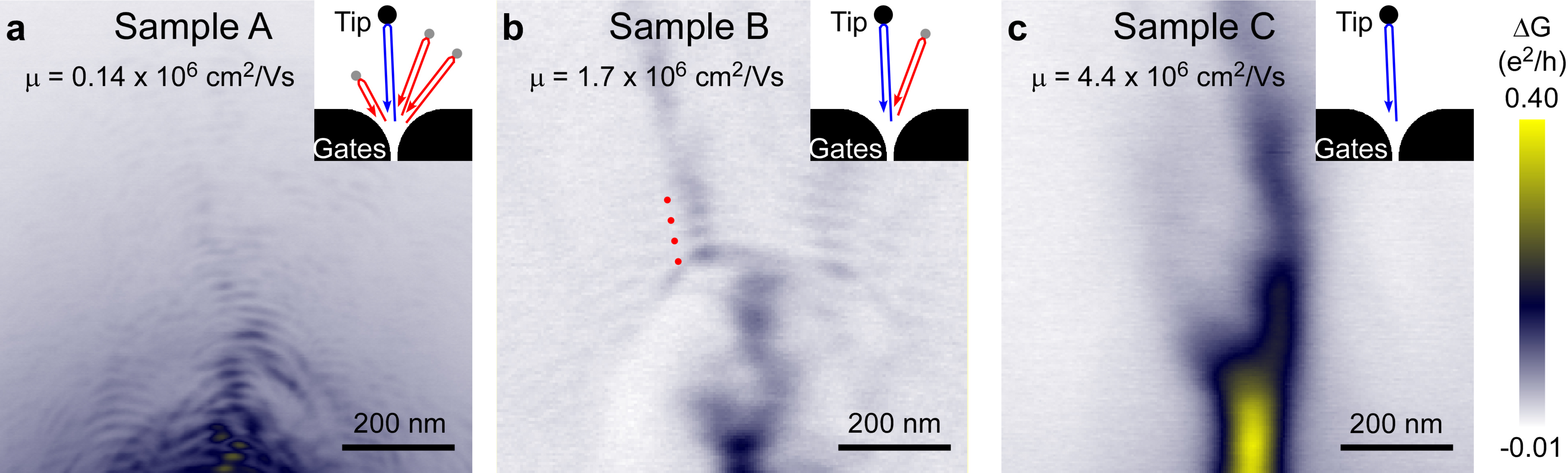}
\end{center}
\caption[Interference fringes in three samples]{\label{fringefig}\footnotesize{\textbf{Interference fringes in samples with different mobilities.}  \textbf{a}-\textbf{c}, Different samples show varying strength of interference fringes, correlated with the samples' mobilities.  The presence of interference fringes more than the thermal length away from the QPC indicates the presence of hard backscattering in the sample due to impurities or defects.  \textbf{a}, Interference fringes in sample A representative of fringes decorating all images of flow in this sample.  \textbf{b}, Electron flow in sample B, which shows interference fringes in some portions and not in others.  Visible interference fringes are rare in sample B; this panel focuses on one of these rare regions.  Fringe spacing is denoted by red circles to the left of electron flow.  \textbf{c}, Electron flow in sample C representative of the absence of interference fringes in this sample.  Insets: Two classes of paths interfere to form fringes (see text): path 1, blue; paths 2, red.  Black regions indicate depletion areas from the gates and the black disc indicates depletion from the tip.  Gray areas indicate hard-scattering sites in the sample, due to impurities close to the 2DEG for example.  The lack of hard-scattering sites (gray dots) in the higher-mobility samples (B and C) suppresses interference fringes in these images.}}
\end{figure}

\newpage

\begin{figure}
\begin{center}
\includegraphics[width=6.5in]{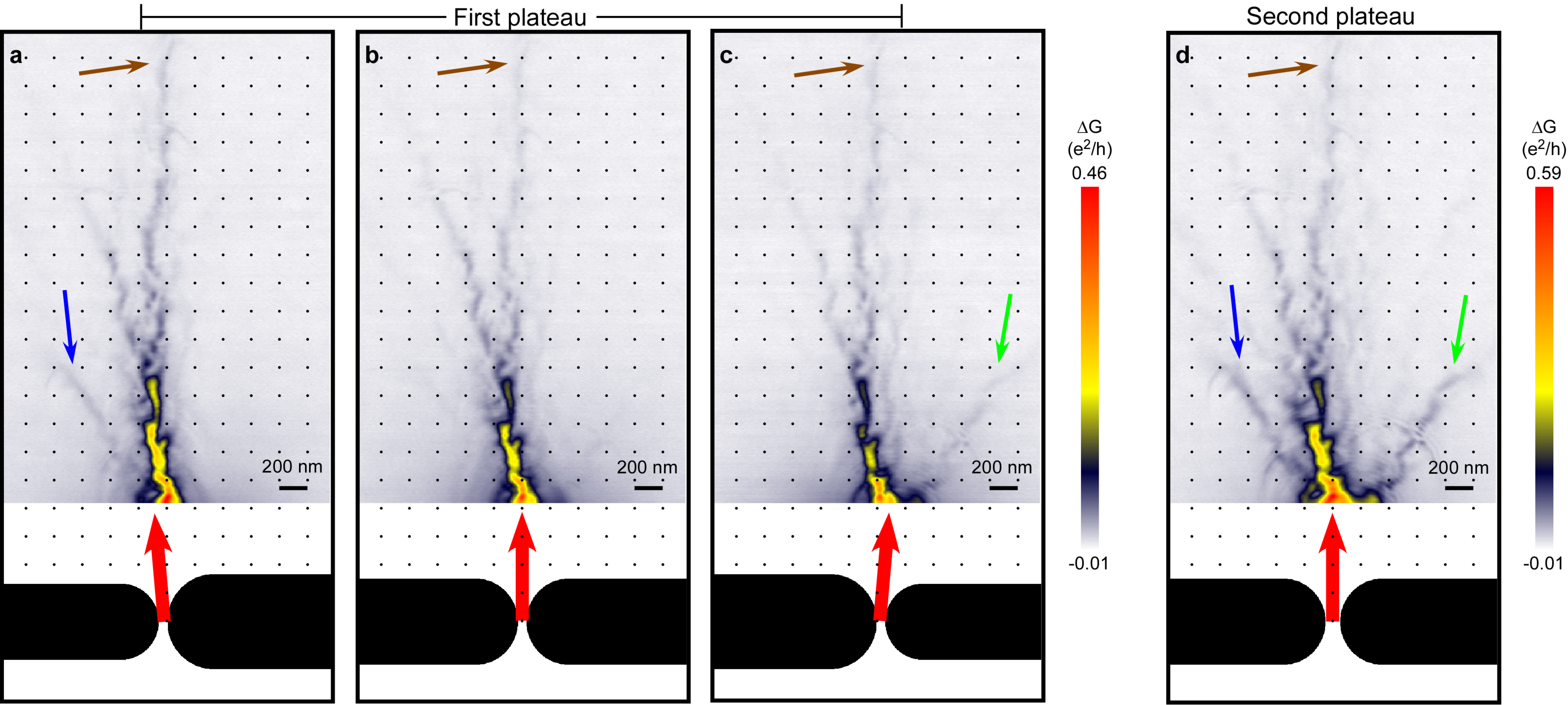}
\end{center}
\caption[Shifting initial conditions]{\label{shiftingfig}\footnotesize{\textbf{SGM images of branch stability in sample B.}  This figure shows the remarkable experimental stability of branched flow.  The reference grid ($200 \ \mathrm{nm}$) denotes the same physical location for precise comparison of electron flow between images.  \textbf{a}-\textbf{c}, Images of electron flow with the QPC open to the first plateau of conductance and electrostatically shifted to the left, not shifted and shifted to the right, respectively.  \textbf{d}, The second plateau of conductance with gates symmetrically energized.  Black regions indicate schematically the depletion regions forming the QPC.  The thick red arrows denote the initial average position and momentum of injected electrons.  On changing the initial conditions of injected electrons, the location of branches does not change, but the current density flowing along each branch does.  In all images, the central branches, one of which is labeled with a brown arrow, are visible.  In \textbf{a} and \textbf{d} the left branches, labeled with a blue arrow, are visible.  In \textbf{c} and \textbf{d} the right branches, labeled with a green arrow, are visible.}}
\end{figure}

\newpage

\begin{figure}
\begin{center}
\includegraphics[width=6.5in]{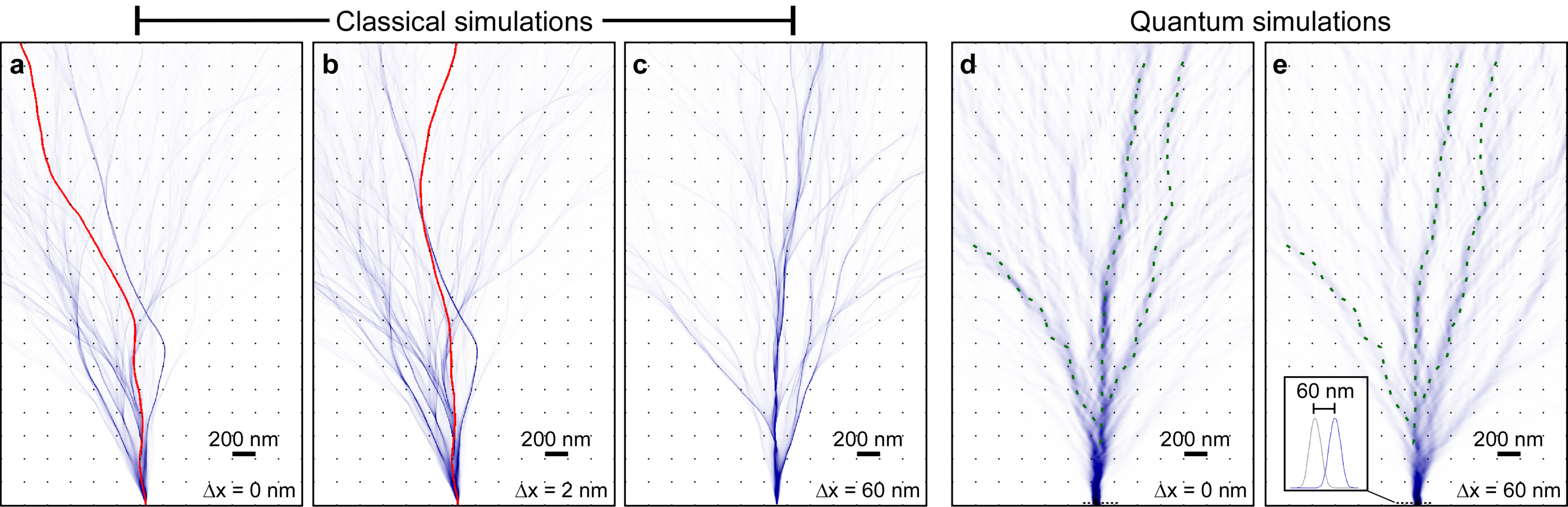}
\end{center}
\caption[Simulation of shifting initial conditions]{\label{simulationfig}\footnotesize{\textbf{Classical and quantum-mechanical numerical investigations of branch stability.}  These figures show simulations of electron flow into a disorder potential chosen to resemble that expected in sample B.  See Supplementary Information, Section I, for log-scale plots.  The reference grid ($200 \ \mathrm{nm}$) denotes the same physical location in every simulation for precise comparison of electron flow between simulations.  \textbf{a}-\textbf{c}, Classical electron trajectories are injected from a point source with $19^{\circ}$ angular spread.  The point source in \textbf{b} is $2 \ \mathrm{nm}$ to the right of that in \textbf{a}, and the point source in \textbf{c} is $60 \ \mathrm{nm}$ to the right of that in \textbf{a}.  The trajectory initially pointing directly up in \textbf{a} and \textbf{b} is highlighted in red.  The slight shift in initial conditions between \textbf{a} and \textbf{b} causes this trajectory (red) to deviate significantly after only $1-2 \ \mathrm{\mu m}$ of travel, but the overall branched flow (blue) formed in \textbf{a} and \textbf{b} still looks surprisingly similar.  However, with a full shift of $60 \ \mathrm{nm}$ (the calculated shift between Fig. 3a and c) the branches formed in \textbf{c} do not resemble those in \textbf{a}.  \textbf{d},\textbf{e},  Quantum-mechanical simulations of flow through an electrostatically shifted QPC into the same disorder potential as \textbf{a}-\textbf{c}.  The injection points in \textbf{d} and \textbf{e} correspond to those in \textbf{a} and \textbf{c}, respectively.  Most branch patterns in \textbf{d} and \textbf{e} remain stable between the two simulations. Some similar branches are denoted with dashed dark green lines.  Inset in \textbf{e}: Lateral profiles of electron probability density ($|\psi|^2$) immediately after injection in \textbf{d} and \textbf{e}.  Electrons are injected $60 \ \mathrm{nm}$ further to the right in \textbf{e} than in \textbf{d} owing to uneven voltage on the gates (for comparison, $\lambda_F = 56 \ \mathrm{nm}$).  The similarity in flow patterns is significant and striking because the overlap between the two injection profiles is minimal.}}
\end{figure}

\end{document}